\documentclass[prl,twocolumn,amssymb,amsmath,floatfix,showpacs,%
superscriptaddress]{revtex4}
\usepackage{graphicx,hyperref}
\newcommand{\vect}[1]{\mathbf{#1}}

\begin{document}
\title{Unambiguous symmetry assignment for the top valence band of ZnO by
magneto-optical studies of the free $A$-exciton state}
\author{Lu Ding}
\affiliation{Department of Physics, Hong Kong University of Science
and Technology, Clear Water Bay, Kowloon, Hong Kong, China}
\author{Chunlei Yang}
\affiliation{Department of Physics, Hong Kong University of Science
and Technology, Clear Water Bay, Kowloon, Hong Kong, China}
\author{Hongtao He}
\affiliation{Department of Physics, Hong Kong University of Science
and Technology, Clear Water Bay, Kowloon, Hong Kong, China}
\author{Fengyi Jiang}
\affiliation{Department of Physics, Nanchang University, Nanchang,
Jiangxi, China}
\author{Jiannong Wang}
\affiliation{Department of Physics, Hong Kong University of Science
and Technology, Clear Water Bay, Kowloon, Hong Kong, China}
\author{Zikang Tang}
\affiliation{Department of Physics, Hong Kong University of Science
and Technology, Clear Water Bay, Kowloon, Hong Kong, China}
\author{Bradley A. Foreman}
\email{phbaf@ust.hk}
\affiliation{Department of Physics, Hong Kong University of Science
and Technology, Clear Water Bay, Kowloon, Hong Kong, China}
\author{Weikun Ge}
\email{phweikun@ust.hk}
\affiliation{Department of Physics, Hong Kong University of Science
and Technology, Clear Water Bay, Kowloon, Hong Kong, China}
% \date{\today}

\begin{abstract}
We studied the circular polarization and angular dependences of the
magneto-photoluminescence spectra of the free $A$-exciton $1S$ state
in wurtzite ZnO at $T$ = 5 K\@.  The circular polarization properties
of the spectra clearly indicate that the top valence band has
$\Gamma_{7}$ symmetry.  The out-of-plane component $B_{\parallel c}$
of the magnetic field, which is parallel to the sample's $c$ axis,
leads to linear Zeeman splitting of both the dipole-allowed
$\Gamma_{5}$ exciton state and the weakly allowed
$\Gamma_{1}$/$\Gamma_{2}$ exciton states.  The in-plane field
$B_{\perp c}$, which is perpendicular to the $c$ axis, increases the
oscillator strength of the weak $\Gamma_{1}$/$\Gamma_{2}$ states by
forming a mixed exciton state.
\end{abstract}


\maketitle

Zinc oxide is a direct wide-gap semiconductor of strong interest for
optoelectronic applications due to its large (60 meV) exciton binding
energy.  Its properties have been studied for many years, with a sharp
increase in activity during the past decade \cite{Ozgur05}.  Despite
its long history, some fundamental properties of ZnO are still not
fully understood.  The valence-band symmetry ordering is especially
controversial.  In most wurtzite semiconductors, the quasidegenerate
$p$-like valence states at $\Gamma$ are split by the crystal-field and
spin-orbit interactions into states of symmetry $\Gamma_9$,
$\Gamma_7$, and $\Gamma_7$ \cite{Birman59b}, in order of decreasing
energy.  However, Thomas \cite{Thomas60} and Hopfield
\cite{Hopfield60}, on the basis of reflectivity studies of fundamental
excitonic transitions, proposed that ZnO has a {\em negative}
spin-orbit splitting, leading to a reversed
$\Gamma_{7}$--$\Gamma_{9}$--$\Gamma_{7}$ ordering.

This reversed ordering is consistent with a wide variety of
experimental data (see Refs.\
\cite{HuHeBa78,Blat82,Fiebig93,WrzeFroh97,WrzeFroh98} for a few
examples), and is also supported by first-principles calculations
\cite{Lask06}.  Nevertheless, some authors have rejected this
interpretation in favor of the conventional
$\Gamma_{9}$--$\Gamma_{7}$--$\Gamma_{7}$ ordering
\cite{Park66,Reynolds99a,Reynolds99b,Gil01,Chichibu03,Chichibu05,%
Adachi05,Gil05}.  Many of the studies supporting reversed ordering did
not directly compare the two possibilities; hence, although these
studies provide cumulative evidence in favor of reversed ordering,
they cannot be said to definitively resolve the controversy.  Some
such studies also used models with a large number of fitting
parameters, leaving open the possibility that other parameter sets
(perhaps consistent with a different ordering) might yield an equally
good fit.

A more direct approach was taken in Refs.\ \cite{Lambrecht02} and
\cite{Rod04}, which used first-principles calculations
\cite{Lambrecht02} and magneto-optical studies of {\em bound} excitons
(BX) \cite{Rod04} to argue that the sign of the hole $g$ factor
deduced from magneto-optical studies of {\em free} excitons (FX) in
Ref.\ \cite{Reynolds99a} is incorrect, and that the top valence band
of ZnO should therefore have $\Gamma_7$ symmetry.  However, as pointed
out by Thomas and Hopfield \cite{ThomHop61}, the hole $g$ factors
derived from studies of BX may, in principle, be entirely different
from the $g$ factors of free holes, due to mixing of the
quasidegenerate valence states by the defect potential.  For this
reason, it is not {\em a priori} obvious that results based on BX are
capable of providing unambiguous evidence for the symmetry of the top
valence band of ZnO.

In view of the simple and well defined nature of FX, we have employed
high-resolution magneto-photo\-lu\-mi\-nes\-cence (PL) of $A$ excitons
to show the valence-band ordering in a more specific and
straightforward way.  A powerful technique, magneto-PL explicitly
reveals the relationship between the fundamental optical transitions
of semiconductors and the optical selection rules that are uniquely
determined by the band structure symmetries.  In this paper,
unambiguous evidence obtained by careful and detailed magneto-PL
measurements is presented to indicate, without any doubt, that the top
valence band of wurtzite ZnO has $\Gamma_7$ symmetry.  This
interpretation is also supported by the polarization dependence of the
Zeeman splitting of neutral-impurity BX.

Free excitons involving the $s$-like $\Gamma_7$ conduction band and
the three valence bands are labeled as $A$, $B$, and $C$ excitons, in
order of increasing exciton energy \cite{Thomas60}.  Depending on the
symmetry assigned to the top valence band, the $A$ excitons have two
possible symmetries:
\begin{equation}
  \Gamma_7 \otimes \Gamma_7 \rightarrow \Gamma_5 \oplus \Gamma_1
  \oplus \Gamma_2 , \qquad \Gamma_7 \otimes \Gamma_9 \rightarrow
  \Gamma_5 \oplus \Gamma_6 . \label{eq:dir_prod}
\end{equation}
Here the doubly degenerate $\Gamma_{5}$ exciton is dipole-allowed for
light polarized normal to the hexagonal $c$ axis $(\vect{E} \perp
\vect{c})$ and the singly degenerate $\Gamma_{1}$ exciton is
dipole-allowed for $\vect{E} \parallel \vect{c}$, whereas the doubly
degenerate $\Gamma_{6}$ exciton and the singly degenerate $\Gamma_{2}$
exciton are dipole-forbidden.

\begin{figure}
  \includegraphics[width=8.5cm,clip]{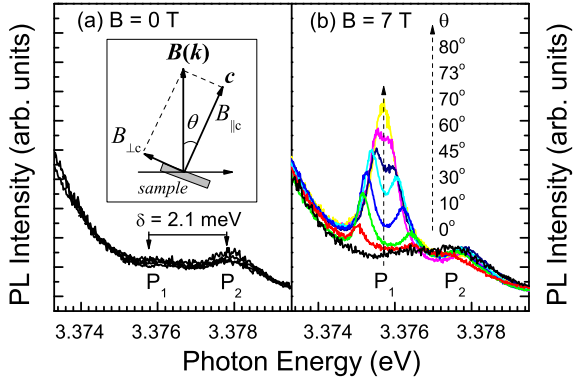}
  \caption{\label{fig:anglePL} (Color online) Angular-dependent PL
    spectra of $\mathrm{FX}_{A}^{n=1}$ at (a) $B = 0$ T and (b) $B =
    7$ T, taken at $T = 5$ K\@. Inset of (a) shows the experimental
    setup.  All spectra are normalized at the higher-energy side of
    $\mathrm{FX}_{A}^{n=1}$.}
\end{figure}

Using a magneto-cryostat with magnetic field $B$ up to 7 T, the
magneto-PL measurements were performed on a 3 $\mu$m thick high
quality ZnO thin film deposited on (0001) sapphire substrate using
metal-organic chemical vapor deposition (MOCVD).  The inset of Fig.\
\ref{fig:anglePL}(a) depicts the magneto-PL experimental setup.
The Faraday configuration ($\vect{k} \parallel \vect{B}$) is applied,
where $\vect{k}$ is the wave vector of the emitted light and $\theta$
is the angle between $\vect{B}$ and the $c$ axis.  $\vect{B}$ can be
decomposed into an out-of-plane component $B_{\parallel c} = B \cos
\theta$ (parallel to the $c$ axis) and an in-plane component $B_{\perp
c} = B \sin \theta$ (perpendicular to the $c$ axis).  In our setup,
different angles $\theta$ were achieved by simply rotating the
$c$ axis.  The incident laser was perpendicular to the magnetic field
for arbitrary $\theta$, except that the backscattering geometry was
used for $\theta = 0$.  The magneto-PL spectra were resolved by a
monochromator (SPEX 1403) with 1800 g/mm double gratings and detected
by a photomultiplier tube (R928).  The spectral resolution of the
system is about 0.1 meV\@.  The circular polarization ($\sigma_{+}$ or
$\sigma_{-}$) of the emitted light was analyzed using a quarter-wave
plate and a linear polarizer.  All the measurements were performed at
5 K to minimize energy shifts induced by thermal fluctuation.

To demonstrate clearly the magnetic field effect, the
angular-dependent zero-field PL as well as magneto-PL spectra of the
$A$-exciton $1S$ state ($\mathrm{FX}_{A}^{n=1}$) are shown for
comparison in Figs.\ \ref{fig:anglePL}(a) and \ref{fig:anglePL}(b),
respectively.  At $B = 0$ T, two resolved fine structures of
$\mathrm{FX}_{A}^{n=1}$ are labeled as $P_1$ (3.3757 eV, weak) and
$P_2$ (3.3778 eV, strong), which correspond to the weakly allowed (or
dipole-forbidden) and dipole-active excitons, respectively [see Fig.\
\ref{fig:anglePL}(a)].  The changes of the peak positions and
intensities are found to be negligible at different $\theta$, which
indicates a weak dependence on the polarization direction of the
incident laser.  Applying a magnetic field of 7 T, rich features are
found with strong angular dependence in the PL spectra [see Fig.\
\ref{fig:anglePL}(b)].  When $\theta = 10^{\circ}$, Zeeman splitting
of $P_1$ is observed with a splitting energy $\Delta E_{P_1}$ as large
as 1.4 meV, whereas $P_2$ remains nearly unchanged.  When $\theta$
increases, $\Delta E_{P_1}$ becomes smaller.  The two split peaks of
$P_1$ finally merge into one at $\theta = 80^{\circ}$.  On the other
hand, the integrated intensity $I_{P_1}$ of $P_1$ increases with
increasing $\theta$ and eventually dominates the
$\mathrm{FX}_{A}^{n=1}$ spectrum.  It is worth noting that there is
almost no change in the magneto-PL spectrum at $\theta = 0^{\circ}$
when $B$ is scanned from 0 T to 7 T, which is due to the weakly
allowed (or dipole-forbidden) nature of $P_1$ at $B_{\perp c} = 0$.
The in-plane magnetic field $B_{\perp c}$ is found to significantly
increase the oscillator strength of $P_1$, which will be explained
below.

We interpret the experimental data using a simple quasi-cubic model
\cite{Hopfield60} in which the crystal-field splitting
$\Delta_{\mathrm{cr}}$ and the spin-orbit splitting
$\Delta_{\mathrm{so}}$ are assumed to satisfy $|\Delta_{\mathrm{so}}|
\ll |\Delta_{\mathrm{cr}}|$ \cite{Thomas60,Hopfield60,Lambrecht02}.
We treat $\Delta_{\mathrm{so}}$ as a perturbation of
$\Delta_{\mathrm{cr}}$, working to first order in the energy and to
zeroth order in the state vector.  If we choose the $z$ and $c$ axes
to be the same, the exciton states formed from the $p_x \pm i p_y$
hole states of $\Gamma_{7}$ symmetry (i.e., the $A$ excitons according
to Thomas and Hopfield) are therefore
\begin{subequations} \label{eq:G512}
  \begin{align}
    | \Gamma_5^{(7)}, \pm \rangle & = | s \pm \rangle | {\pm 1}, \mp
      \rangle & ( g_{\mathrm{exc}} & = g_{h}^{\parallel} + g_{e} ) ,
      \label{eq:G57} \\ | \Gamma_{1 \oplus 2}, \pm \rangle & = | s \mp
      \rangle | {\pm 1}, \mp \rangle & ( g_{\mathrm{exc}} & =
      g_{h}^{\parallel} - g_{e} ) . \label{eq:G12}
  \end{align}
\end{subequations}
Here $|s+\rangle |m,-\rangle$ is the tensor product of a spin-up $s$
electron and a spin-down $p$ hole whose $z$ component of orbital
angular momentum is $m$.  The $\pm$ label of the exciton states is
taken from the sign of $m$ (note that for $\Gamma_{5}$, $m$ is also
the $z$ component of the total exciton angular momentum).  In Eq.\
(\ref{eq:G12}), the contribution of $\Delta_{\mathrm{so}}$ to the
short-range exchange interaction is neglected, so that $\Gamma_{1}$
and $\Gamma_{2}$ form an approximately doubly-degenerate reducible
representation \cite{Thomas60,Hopfield60,Lambrecht02,Rod04} denoted
$\Gamma_{1 \oplus 2}$.  A small field $B_{\parallel c}$ produces a
linear Zeeman splitting with the given exciton effective $g$ factors
$g_{\mathrm{exc}}$, in which $g_{e}$ is the (nearly) isotropic
electron $g$ factor and $g_{h}^{\parallel}$ is the hole $g$ factor
parallel to the $c$ axis \cite{Rod04}.  In the simple model of Ref.\
\cite{Lambrecht02} we have $g_{h}^{\parallel} = 2 K - g_0$, where $K =
-(3 \kappa + 1)$ is the magnetic Luttinger parameter and $g_0 = 2$ is
the $g$ factor of a free hole.  The states in Eq.\ (\ref{eq:G12}) are
dipole-forbidden when $B_{\perp c} = 0$, but they become
dipole-allowed when $B_{\perp c} \ne 0$ due to mixing with $|
\Gamma_5^{(7)}, \pm \rangle$ caused by $g_{e}$.

Likewise, the exciton states formed from the $p_x \pm i p_y$ hole
states of $\Gamma_{9}$ symmetry (i.e., the $B$ excitons according to
Thomas and Hopfield) are given by
\begin{subequations} \label{eq:G56}
  \begin{align}
    | \Gamma_5^{(9)}, \pm \rangle & = | s \mp \rangle | {\pm 1}, \pm
      \rangle & ( g_{\mathrm{exc}} & = g_{h}^{\parallel} - g_{e} ) ,
      \label{eq:G59} \\ | \Gamma_{6}, \pm \rangle & = | s \pm
      \rangle | {\pm 1}, \pm \rangle & ( g_{\mathrm{exc}} & =
      g_{h}^{\parallel} + g_{e} ) , \label{eq:G6}
  \end{align}
\end{subequations}
in which $g_{h}^{\parallel} = 2 K + g_0$.  Just as for $| \Gamma_{1
\oplus 2}, \pm \rangle$, the states $| \Gamma_{6}, \pm \rangle$ are
dipole-forbidden when $B_{\perp c} = 0$, but become dipole-allowed
when $B_{\perp c} \ne 0$ due to $g_{e}$-induced mixing with $|
\Gamma_5^{(9)}, \pm \rangle$.

The above model is crude, but it has the advantage of explaining the
main features of the experiment in a simple way.  We have also
considered a more complicated 12-dimensional $1S$-exciton Hamiltonian
\cite{Lambrecht02} that includes a full treatment of spin-orbit
coupling and the short- and long-range exchange interactions, but the
results were qualitatively the same as those obtained from the simple
model defined above (so far as the description of the present
experimental data is concerned).  Therefore, we discuss only the
simple model in this paper.

\begin{figure}
  \includegraphics[width=8.5cm,clip]{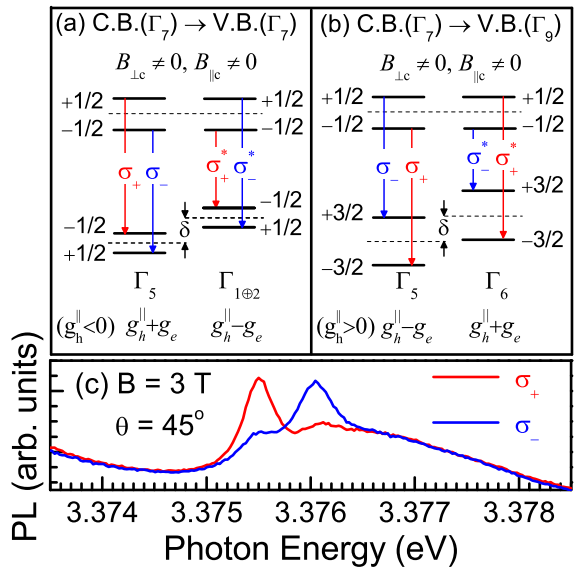}
  \caption{\label{fig:circPL} (Color online) Schematic representations
    of energy levels of $A$-exciton transitions involving holes of (a)
    $\Gamma_{7}$ symmetry and (b) $\Gamma_{9}$ symmetry.  (c) shows the
    circular polarization dependence of the magneto-PL of
    $\mathrm{FX}_{A}^{n=1}$.}
\end{figure}

In Fig.\ \ref{fig:circPL}, we sketch two sets of optically allowed
exciton transitions in a magnetic field with arbitrary $\theta$ (so
that $B_{\parallel c}$ and $B_{\perp c}$ are both nonzero) for the
ground-state free excitons involving a hole of either (a) $\Gamma_{7}$
symmetry or (b) $\Gamma_{9}$ symmetry.
Here $\delta$ is the zero-field exchange splitting between
$\Gamma_{5}$ and $\Gamma_{1 \oplus 2}$ states in case (a) or between
$\Gamma_{5}$ and $\Gamma_{6}$ states in case (b).  The labels $\pm
\frac12$ and $\pm \frac32$ in Fig.\ \ref{fig:circPL} refer to the $z$
component of total angular momentum for conduction and valence
electrons.  The notation $\sigma_{\pm}^{*}$ indicates that these
transitions are dipole-forbidden when $B_{\perp c} = 0$, but emit
photons with $\sigma_{\pm}$ polarization when $B_{\perp c} \ne 0$.
The sign of $g_{h}^{\parallel}$ would have to be negative for
$\Gamma_{7}$ and positive for $\Gamma_{9}$ in order to agree with the
experimental observation that the Zeeman splitting of the weakly
allowed or dipole-forbidden states is much larger than that of the
dipole-active states.

Based on the information in Eqs.\ (\ref{eq:G512}) and (\ref{eq:G56})
and the energy diagrams in Figs.\ \ref{fig:circPL}(a) and
\ref{fig:circPL}(b), it is evident that the symmetry of the top
valence band can be identified by measuring the polarization of the
weakly allowed or dipole-forbidden exciton states under an applied
magnetic field.  For exciton transitions involving a $\Gamma_{9}$ hole
and a $\Gamma_{7}$ electron, one would expect the originally
dipole-forbidden states ($\Gamma_{6}$ excitons) to split, with the
lower-energy peak showing $\sigma_{-}$ polarization. However, if both
the electron and hole have $\Gamma_{7}$ symmetry, the originally
weakly allowed $\Gamma_{1 \oplus 2}$ excitons will show $\sigma_{+}$
polarization for the lower-energy peak.  Figure \ref{fig:circPL}(c)
presents the polarization dependence of the magneto-PL of the
$A$-exciton state with $B = 3$ T and $\theta = 45^{\circ}$.  This
clearly indicates that the lower-energy peak of $P_1$ has $\sigma_{+}$
polarization, which unambiguously demonstrates that the hole in the
$A$-exciton $1S$ state (or the top valence band) in wurtzite ZnO has
$\Gamma_{7}$ symmetry.  The experimentally determined zero-field
exchange splitting $\delta$ is 2.1 meV, which is in good agreement
with Refs.\ \cite{Ozgur05}, \cite{Thomas60}, and \cite{HopThom65}.

\begin{figure}
  \includegraphics[width=8.5cm,clip]{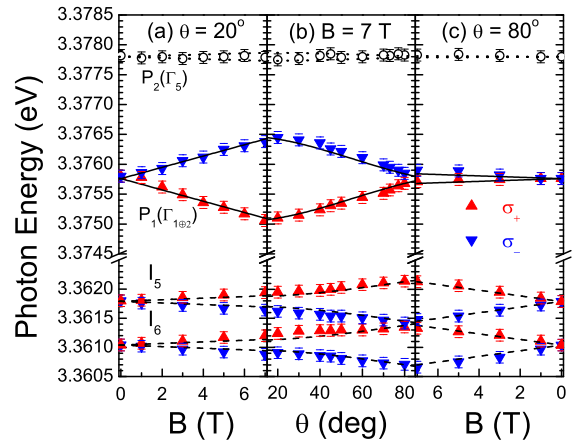}
  \caption{\label{fig:Zeeman} (Color online) The magnetic-field and
    angular dependences of the peak energies of $A$ excitons ($P_1$
    and $P_2$) and BXs ($I_5$ and $I_6$), as described in the
    text.}
\end{figure}

To get more information on the electron and hole $g$ factors, the
magnetic field dependences of the transition energies of $P_1$ and
$P_2$ are summarized in Figs.\ \ref{fig:Zeeman}(a) and
\ref{fig:Zeeman}(c) for $\theta = 20^{\circ}$ and $\theta =
80^{\circ}$, respectively.
Figure \ref{fig:Zeeman}(b) shows the $\theta$ dependence of $P_1$ and
$P_2$ at $B = 7$ T\@.  In the Zeeman splitting of $P_1$ and $P_2$,
$B_{\parallel c}$ lifts the degeneracy of the $P_1$ ($\Gamma_{1 \oplus
2}$) states or the doublet $P_2$ ($\Gamma_{5}$) state.  The energy
splitting of $P_1$ ($\Gamma_{1 \oplus 2}$) is fitted using $E_{P_1
\pm} = E_{P_1} \pm \frac12 (g_{h}^{\parallel} - g_{e}) \mu_B
B_{\parallel c}$, where $\mu_B$ is the Bohr magneton and $E_{P_1} =
3.37576$ eV is the zero-field transition energy of $P_1$ ($\Gamma_{1
\oplus 2}$).  Using $g_{e} = 1.95$ \cite{Reynolds65}, the hole $g$
factor obtained from the fitting (see solid curves in Fig.\
\ref{fig:Zeeman}) is $g_{h}^{\parallel} = -1.6$, which agrees well
with the values obtained in Refs.\ \cite{HuHeBa78} and \cite{Blat82}
(but with a different convention for the sign of $g_{h}^{\parallel}$).
The fact that the Zeeman splitting for $P_2$ ($\Gamma_{5}$) could not
be resolved (see black dots in Fig.\ \ref{fig:Zeeman}) indicates the
nearly equal absolute values of $g_{e}$ and $g_{h}^{\parallel}$.  The
dotted curves for $P_2$ are plotted according to $E_{P_2 \pm} =
E_{P_2} \pm \frac12 (g_{h}^{\parallel} + g_{e}) \mu_B B_{\parallel
c}$, employing $g_{e} = 1.95$ and $g_{h}^{\parallel} = -1.6$.

In addition, the Zeeman splitting of BXs $I_5$ and $I_6$
\cite{Meyer04} has also been observed, and the transition energies are
shown in Fig.\ \ref{fig:Zeeman}.  The circular polarization
dependences indicate that $I_5$ and $I_6$ are excitons bound to
neutral impurity centers with $A$ holes involved \cite{Rod04}.  The
dashed lines are fitted results given by $\pm \frac12 \mu_B B (g_{e} +
g_{h})$ and $g_{h} = g_{h}^{\parallel} \sqrt{\cos^2 \theta +
(g_{h}^{\perp} / g_{h}^{\parallel})^2 \sin^2 \theta}$, where $g_{e} =
1.95$, $g_{h}^{\parallel} = -1.6$, and $g_{h}^{\perp} = 0.11$.  The
equality of the fitted FX and BX values of $g_{h}^{\parallel}$
provides {\em ex post facto} support for the conclusions of Ref.\
\cite{Rod04} (although, as noted in the introduction, such similarity
cannot be assumed to hold in general).

\begin{figure}
  \includegraphics[width=8.5cm,clip]{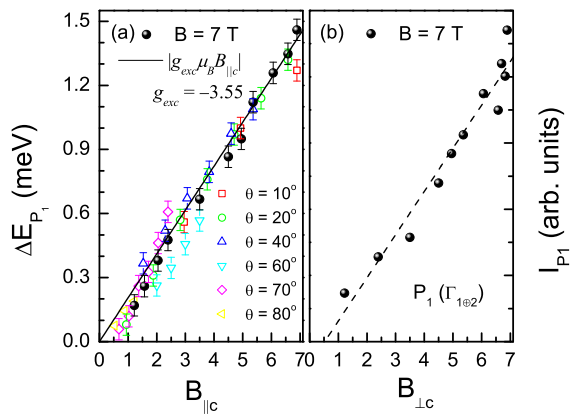}
  \caption{\label{fig:linear} (Color online) (a) $B_{\parallel c}$
    dependence of the Zeeman splitting $\Delta E_{P_1}$ of $P_1$
    ($\Gamma_{1 \oplus 2}$). (b) $B_{\perp c}$ dependence of the
    intensity $I_{P_1}$ of $P_1$ ($\Gamma_{1 \oplus 2}$) (solid
    dots). The dashed line is a guide for the eyes.}
\end{figure}

The different contributions of the in-plane and out-of-plane magnetic
field to the magneto-PL spectra are shown more specifically in Fig.\
\ref{fig:linear}.  
The left panel [Fig.\ \ref{fig:linear}(a)] shows the measured
$B_{\parallel c}$ dependence of the Zeeman splitting $\Delta E_{P_1}$
of $P_1$ ($\Gamma_{1 \oplus 2}$).  The data taken at $B = 7$ T for
different $\theta$ (solid black dots) and those taken at fixed
$\theta$ for different $B$ (hollow colored dots) fall onto the same
line plotted using the equation $\Delta E_{P_1} = |g_{\mathrm{exc}}
\mu_B B_{\parallel c}|$ with $g_{\mathrm{exc}} = g_{h}^{\parallel} -
g_{e} = -3.55$.  The zero-field splitting of the $\Gamma_{1}$ and
$\Gamma_{2}$ states is zero as expected.  This good linear
relationship between $\Delta E_{P_1}$ and $B_{\parallel c}$ reveals
that the splitting of the $A$-exciton states depends on the
out-of-plane field instead of the total magnetic field, which can be
well explained by the $\Gamma_{2}$ symmetry of the out-of-plane field
that mixes $\Gamma_{1}$ only with $\Gamma_{2}$ states
\cite{Hopfield60}.  Figure \ref{fig:linear}(b) shows that the
intensity $I_{P_1}$ of $P_1$ increases monotonically with increasing
$B_{\perp c}$.  The transition probability of the originally weakly
allowed $\Gamma_{1}$/$\Gamma_{2}$ excitons increases significantly due
to mixing with $\Gamma_{5}$ excitons.

In summary, angular-resolved magneto-PL measurements were applied to a
high quality ZnO thin film with circular polarization analysis.  The
top valence band of wurtzite ZnO was found to have $\Gamma_{7}$
symmetry with no ambiguity by directly examining the polarization of
the $A$-exciton emission.  The out-of-plane component $B_{\parallel c}$
of the magnetic field was found to be responsible for the linear
Zeeman splitting of the $\Gamma_{5}$ and $\Gamma_{1}$/$\Gamma_{2}$
states.  The in-plane magnetic field $B_{\perp c}$ increases the
oscillator strength of the originally weakly allowed
$\Gamma_{1}$/$\Gamma_{2}$ states by mixing with $\Gamma_{5}$ states.
The hole effective $g$ factor was found to be negative and has the value
$-1.6$.

\begin{acknowledgments}
Thanks are due to Professor G. Q. Hai of the Universidade de S\~ao
Paulo and Professor Y. Q. Wang of the Institute of Solid State
Physics, Chinese Academy of Sciences for encouraging discussions. This
work is funded by the Hong Kong University of Science and Technology
via grants no. DAG04/05.SC24 and DAG05/06.SC30.
\end{acknowledgments}

% \bibliography{main}

\end{document}